\documentclass[manuscript, nonacm] {acmart} 
\AtBeginDocument{%
  \providecommand\BibTeX{{%
    \normalfont B\kern-0.5em{\scshape i\kern-0.25em b}\kern-0.8em\TeX}}}

\setcopyright{acmlicensed}
\copyrightyear{2018}
\acmYear{2018}
\acmDOI{XXXXXXX.XXXXXXX}

\acmConference[CI'2024]{Make sure to enter the correct
  conference title from your rights confirmation emai}{June 26--29,
  2024}{Boston, MA}
\acmBooktitle{Boston '24: ACM Collective Intelligence,
 June 26--29, 2024, Boston, MA} 
\acmISBN{978-1-4503-XXXX-X/24/06}

\title{GigSense: An LLM-Infused Tool for Workers' Collective Intelligence} 
\author{Kashif Imteyaz}
\affiliation{%
  \institution{Northeastern University}
  \city{Boston}
  \state{MA}
  \country{USA}}
\email{imteyaz.k@northeastern.edu}

\author{Claudia	Flores-Saviaga}
\affiliation{%
  \institution{Northeastern University}
  \city{Boston}
  \state{MA}
  \country{USA}
\email{floressaviaga.c@northeastern.edu}
}

\author{Saiph Savage}
\affiliation{%
  \institution{Northeastern University}
  \city{Boston}
  \state{MA}
  \country{USA}
\email{s.savage@northeastern.edu}
}

\begin{document}

\begin{teaserfigure}
\centering
  \includegraphics[width=1\textwidth, alt= Diagram showing an overview of GigSense's functionality. It has an intelligent agent that organizes and summarizes workers' problems. Workers can analyze collective issues by zooming in and out. GigSense also provides a collaborative space for brainstorming and selecting solutions. The intelligent agent additionally offers AI-enhanced solutions to support workers' brainstorming and planning]{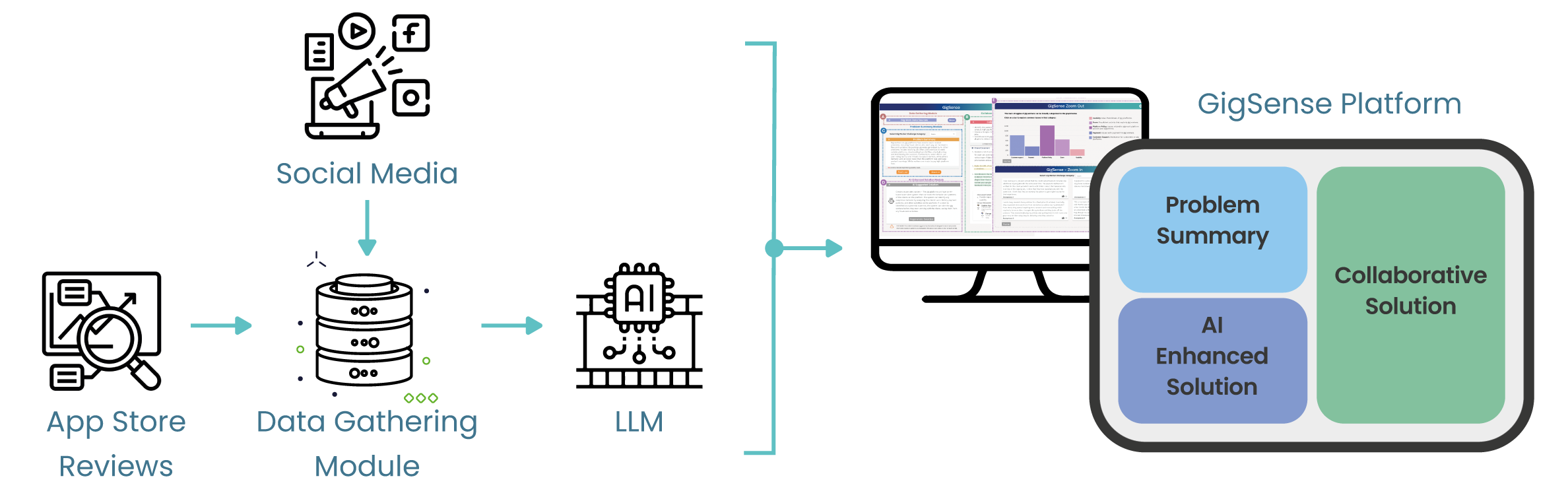}
  \caption{Overview of GigSense's functionality: GigSense first features an intelligent agent that automatically organizes and summarizes workers' problems. Within short timeframes, workers can zoom in and out to analyze and understand their collective issues. The tool also offers a collaborative space for brainstorming and selecting optimal solutions to address their problems. Moreover, GigSense's intelligent agent provides AI-enhanced solutions to further support workers' brainstorming and planning endeavors.}
  \label{fig:plansInterface}
\end{teaserfigure}
\maketitle

\section*{\bf Abstract}
Collective intelligence among gig workers yields considerable advantages, including improved information exchange, deeper social bonds, and stronger advocacy for better labor conditions. Especially as it enables workers to collaboratively pinpoint shared challenges and devise optimal strategies for addressing these issues.  However, enabling collective intelligence remains challenging, as existing tools often overestimate gig workers' available time and uniformity in analytical reasoning. To overcome this, we introduce GigSense, a tool that leverages large language models alongside theories of collective intelligence and sensemaking. GigSense enables gig workers to rapidly understand and address shared challenges effectively, irrespective of their diverse backgrounds. Our user study showed that GigSense users outperformed those using a control interface in problem identification and generated solutions more quickly and of higher quality, with better usability experiences reported. GigSense not only empowers gig workers but also opens up new possibilities for supporting workers more broadly, demonstrating the potential of large language model interfaces to enhance collective intelligence efforts in the evolving workplace.

\section{\bf Introduction}
Harnessing collective intelligence among gig workers can significantly enhance their ability to improve labor conditions \cite{chen2023worker,qadri2023worker,grohmann2023worker,piven2012poor}. This collaborative approach enables workers to collectively deliberate on their challenges, devise strategic solutions, and collaboratively implement their action plans to tackle these issues effectively\cite{globalanalysisofdigitalworkprotest,UberSuccess}.

However, despite occasional successes, collective intelligence among gig workers is uncommon, leaving many issues unresolved \cite{globalanalysisofdigitalworkprotest}. This scarcity largely stems from the lack of technologies designed for facilitating collective problem-solving among gig workers \cite{johnston2020labour,shaw2014computer,zhang2014wedo,gray2019ghost,lehdonvirta2022cloud}. Platforms like Dynamo and Coworker.org, which allow for sharing, prioritizing, and solving issues through upvote-based lists, fail to provide a comprehensive view of problems and solutions \cite{salehi2015we,demaria2016future}. Such list-based interfaces limit the exploration of issues from multiple angles and understanding the full scope of potential solutions \cite{li2023dimensions,10.1145/3359310,10.1145/3533406.3533424,mattsson2015sense,sentido,weick1995sensemaking}, often leading workers to prioritize less critical concerns \cite{wood2019platform}. This impedes meaningful progress in addressing the fundamental challenges of gig work, especially during the initial phases of collective intelligence focused on recognizing shared problems and formulating solutions \cite{boomorbane}. 

Another hurdle is that the diverse and significant time commitments of gig workers may limit their ability to contribute to collective intelligence, underscoring the importance of developing inclusive tools that can accommodate workers' varying schedules, reducing the barriers to collective intelligence and enabling easier participation in collaboratively tackling challenges \cite{packham2008active,pantea2013changing}.
To address these challenges, we introduce GigSense, a novel platform for fostering collective intelligence among gig workers by enabling collaborative problem-solving. By integrating Sense-Making Theory and leveraging the capabilities of large-language models (LLMs), GigSense features an LLM-enhanced interface that allows workers to deeply analyze their challenges and solutions from multiple perspectives \cite{sentido}. Unlike existing platforms that offer a simple list-based view of issues \cite{salehi2015we}, GigSense enables a detailed exploration of problems, offering both a close-up and a broad overview to understand workplace dynamics better. Additionally, GigSense uses LLMs to facilitate collective brainstorming, helping workers develop solutions together. Our user study indicates that GigSense enhances the sensemaking process for gig workers, accelerating their ability to identify a broader range of problems and generate more feasible, diverse solutions. As a significant advancement in fostering collective intelligence among gig workers, GigSense addresses the unique challenges of this workforce, which typically operates in isolation without a common platform for problem identification and solution generation \cite{10.1145/3533406.3533424,yao2021together}. GigSense bridges this gap by offering a communal space for gig workers to collectively understand and address their issue. Leveraging shared digital resources to build community \cite{sawyer2014digital,barley1986technology}, GigSense emerges as a vital tool for a fragmented workforce, potentially catalyzing the formation of supportive gig worker communities \cite{tajfel1979integrative,cohen2003collective,van2008toward,tajfel2004social}. Beyond organization, it could reveal systemic issues, encouraging workers to see problems like client disputes and payment delays as collective challenges. This recognition can promote solidarity, crucial for collective intelligence, and empower workers to develop effective negotiation strategies, potentially improving their working conditions.

In this paper, we highlight two main contributions: (1) a system that integrates LLMs with an interactive interface to support gig workers in problem-solving and kickstart collective intelligence among workers, and (2) a study showing GigSense not only speeds up problem-solving but also improves the quality and variety of worker-generated solutions. We discuss the influence of LLMs on sensemaking and collective intelligence and its implications for workers, proposing directions for future research.

\section{\bf Related Work}
\subsubsection*{\bf Gig Work and Platform} 
``Gig" or platform-based work is a significant trend in the labor market, driven by the demand for flexibility from both employers and workers \cite{watson2021looking}, and facilitated by digital technology \cite{mandl2015new,de2016rise,johnston2018organizing}. While offering economic benefits to disadvantaged groups, gig work also presents challenges like unstable schedules, income variability, and uncertain long-term job security \cite{peck2012politicizing,de2016rise}. Workers have engaged in collective actions (e.g., negotiations, strikes, unionization) to improve conditions, yet face obstacles due to platform constraints, geographic dispersion, and a lack of community and common interest, making organizing difficult \cite{gray2019ghost,johnston2018organizing,sutherland2017gig,yao2021together,webster2016microworkers}. Despite these challenges, some success has been noted in collective efforts \cite{UberSuccess}, but tools to support organizing are limited \cite{johnston2018organizing}. This paper proposes a tool to aid gig workers in understanding their challenges and initiating collective intelligence.

\subsubsection*{\bf Sensemaking and Collective Intelligence} Comprehending the challenges faced by gig workers can be perceived as an act of sensemaking, involving the collection and analysis of diverse and unstructured data to reach a conclusion. 
Pirolli and Card \cite{sentido} define sensemaking as a series of iterative steps. For instance, it starts with the initial gathering of relevant data ({\it ``Step: Search and Filter''}), akin to brainstorming gig workers' problems. Subsequently, it involves extracting valuable information ({\it ``Step: Read and Extract''}), akin to selecting the most pertinent issues. Further, it encompasses summarizing and schematizing the information ({\it ``Step: Schematize''}), akin to the manual procedure of condensing and structuring of the brainstormed ideas. Then, it involves generating hypotheses from various perspectives ({\it ``Step: Build Case''}), resembling the development of viable solutions. Lastly, it culminates in decision-making to determine the best solution ({\it ``Step: Tell Story'')}. Smith \cite{smith1994collective} defines collective intelligence as a group of individuals working together on tasks, where the collective group itself demonstrates coherence, intelligence, and constant improvement, enabling more effective mobilization of skills than any single individual working independently. Significant research explores collaborative sensemaking tools for domains like literature review \cite{zhang2008citesense}, web search \cite{paul2009cosense}, organizing academic literature \cite{reimer2011turning}, solving mysteries \cite{li2018crowdia}, and tackling disinformation \cite{flores2022datavoidant}. We introduce GigSense to aid gig workers' collaborative sensemaking, automating parts of the process pipeline. We also leverage Large Language Models to enhance collective intelligence.

\subsubsection*{\bf LLMs for Idea Generation.}
In recent years, significant advancements in large language models (LLMs) have positioned them as a promising tool for facilitating a diverse array of writing tasks, such as story generation \cite{mirowski2023co,chung2022talebrush,yuan2022wordcraft,alabdulkarim2021automatic}, academic writing \cite{gero2022sparks}, question-answering \cite{brown2020language}, idea generation \cite{girotra2023ideas} and information synthesis \cite{liu2023selenite,palani2023relatedly}. However, despite their impressive utility, LLMs have faced criticism for generating text that, while appearing logically and grammatically coherent, may contain factual inaccuracies or lack meaningful coherence (referred to as hallucinations) \cite{ji2023survey}.  Nevertheless, researchers suggest that this weakness can be reframed as a strength \cite{girotra2023ideas}, and have started to use LLMs for idea suggestions\cite{yuan2022wordcraft}. In creative writing and problem-solving, possessing a range of idea quality and quantity holds more value than maintaining unwavering consistency \cite{girotra2010idea}. To achieve a wide range of ideas with varying levels of quality, most ideation research advises generating numerous ideas first and delaying their evaluation \cite{girotra2010idea}. LLMs are designed to do exactly this— quickly generate many somewhat plausible solutions \cite{girotra2023ideas}. 
In this research, we align with the principle that an unrestrained influx of ideas can often pave the way for innovative solutions. Therefore, we harness the capabilities of LLMs by integrating them into Gigsense to assist gig workers in rapidly producing multiple reasonably viable ideas for their most pressing challenges.

\subsubsection*{\bf Interfaces for Visualizing Collective Problems and Solutions}.
Other research that has inspired our work includes interfaces and systems designed to assist individuals in visualizing both problems and solutions. For example, MacNeil et al. \cite{macneil2021probmap}, presented a design gallery for visualizing problems and the stakeholders involved in those problems. Similarly, Huang et al. \cite{huang2023designnet} introduced a novel system to aid designers in comprehending various issues within a space and exploring diverse solutions. Their research focused on mitigating design fixation by encouraging a broader consideration of context and essential relationships during the design process.  Siangliulue et al.'s system \cite{siangliulue2016ideahound} innovatively merged crowdsourcing with machine learning to create a semantic solution space model, whihelped to foster user creativity and diversity of ideas. Our research is inspired by the principles of these systems, concentrating on assisting gig workers, who may lack expertise in design and problem-solving, in exploring and comprehending various aspects of problems and having support to provide solutions.
\\

\section{\bf GigSense}
\begin{figure}[h]
\centering
  \includegraphics[width=.8\textwidth, alt= Diagram of GigSense's problem-solving interface showing four modules: A) Data Gathering Module; B) Collaborative Solution Module where workers can collaborate on solutions to specific problems; C) Problem Summary Module where workers get summaries of specific problems to aid sensemaking; D) AI-Enhanced Solution Module where workers get solution suggestions from large language models.]{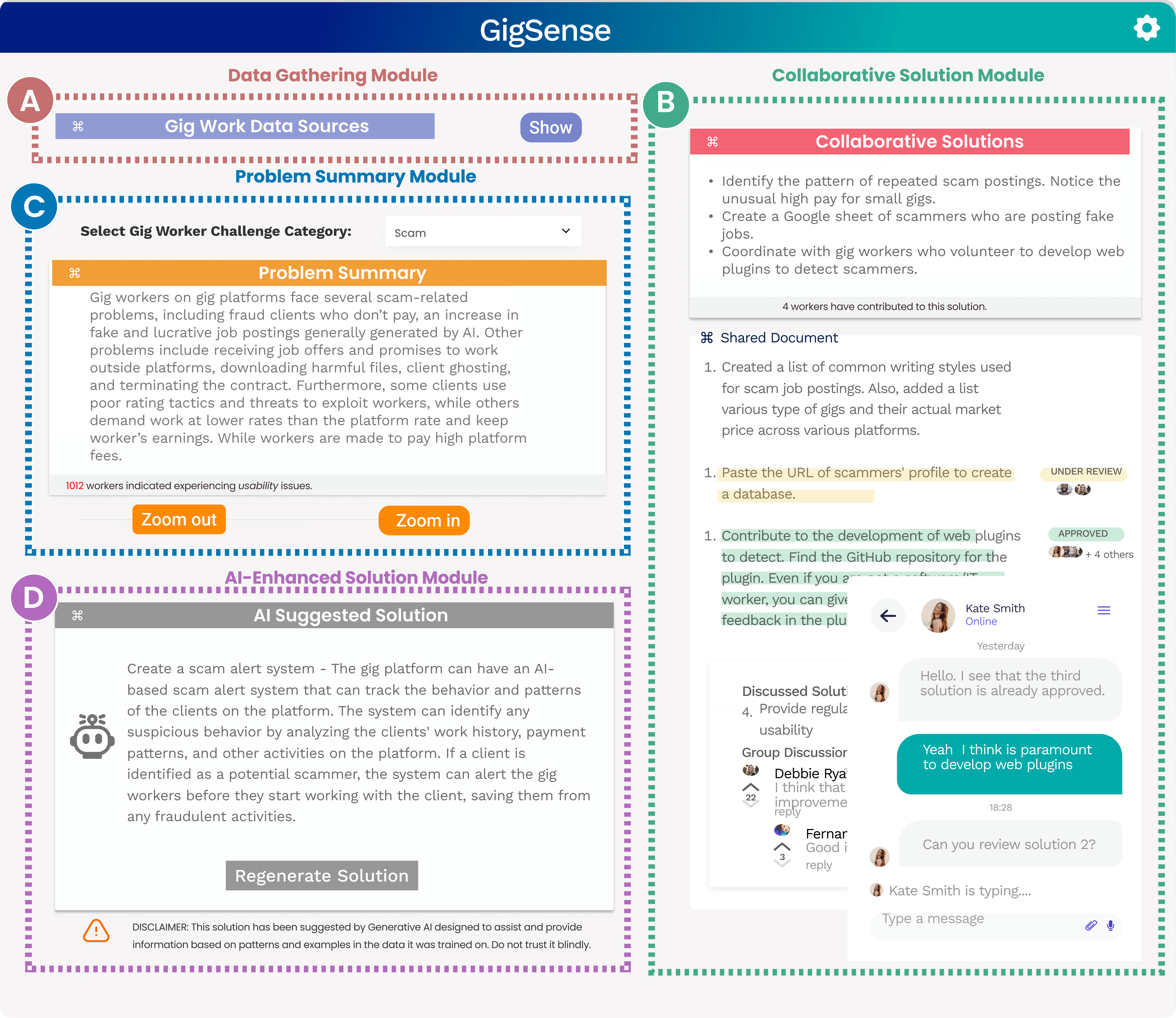}
  \caption{{\small Overview of GigSense's problem-solving interface with the: A) ``Data Gathering Module'';  B) ``Collaborative Solution Module'' where workers can collaborate to create solutions to specific problems; C) ``Problem Summary Module'' where workers obtain summaries of specific problems to help their sensemaking; and D) ``AI-Enhanced Solution Module'' where workers obtain solution suggestions from LLMs.}}
  \label{fig:gigsense}

  \vspace{-0.5cm}
\end{figure}

\subsection{\bf Description of GigSense}

GigSense is an AI-enhanced platform designed to assist gig workers in the early phases of sensemaking and collective intelligence, which includes problem identification and solution proposal \cite{shaw2014computer}. Our platform thus focuses on aiding workers in their sensemaking process (identifying collective issues and potential solutions). For this purpose, GigSense incorporates Pirolli and Card's sensemaking theories and collective intelligence theories into various modules within its system \cite{shaw2014computer,weick1995sensemaking,CIBook}. Fig \ref{fig:gigsense} presents GigSense's interface and its modules. These modules are designed to automate the sensemaking process, which unfolds seamlessly within the entirety of GigSense's interface, with workers playing a crucial role in steering this process.

GigSense's modules are based on previous designs \cite{flores2022datavoidant,li2018crowdia}, but uniquely integrate human-AI collaboration to address specific needs and contexts of gig workers (see Fig \ref{fig:gigsense}.C-D.) GigSense recognizes the time constraints faced by gig workers when engaging in collective intelligence and hence provides interfaces that facilitate rapid sensemaking of problems and proposals of solutions. For this purpose, it utilizes Large Language Models (LLMs) to provide concise summaries of issues found on gig platforms, enabling workers to quickly understand prevalent problems. By leveraging the OpenAI GPT-4 API \cite{openaiOpenAIPlatform}, it categorizes data and generates these summaries, aiding workers in gaining insights and brainstorming solutions. Our system also offers an interactive interface, allowing users to ``zoom in'' and ``zoom out'' and explore specific discussions related to a problem. While LLM supports the creation of problem summaries, GigSense's interface design empowers workers to independently assess the information. It also provides a collaborative workspace for workers to propose solutions to identified problems. It uses LLMs to inspire workers with potential solutions but prioritizes human-generated ideas over AI suggestions to emphasize the value of human input. Notice that GigSense's design, focusing on quick sensemaking, contrasts with traditional extended analysis but offers notable advantages. It enables time-constrained individuals like gig workers with multiple jobs, parents, and students to participate, enhancing diversity and inclusivity. This approach makes sensemaking more accessible, as it requires less long-term commitment \cite{etz2018rapid}. A faster process encourages broader participation and engagement, as people are more likely to contribute when their time investment is minimal yet impactful \cite{sandberg2015making}. This approach can be especially beneficial for future engagements around collective intelligence, such as implementing solutions, where sustained participant involvement is crucial \cite{etz2018rapid}. Additionally, rapid sensemaking can foster concise and focused information processing, reducing data overload and maintaining participant interest, which can wane in longer processes \cite{mulgan2018big,sandberg2015making}. Next, we describe each GigSense module in detail:

\begin{figure}[h]
\centering
  \includegraphics[width=1\textwidth, alt= Diagram showing GigSense's zoom-out and zoom-in views accessible via interface buttons. Zoom-out shows an overall view of the problems. Zoom-in shows a concise summary of specific problems being analyzed and relevant posts.]{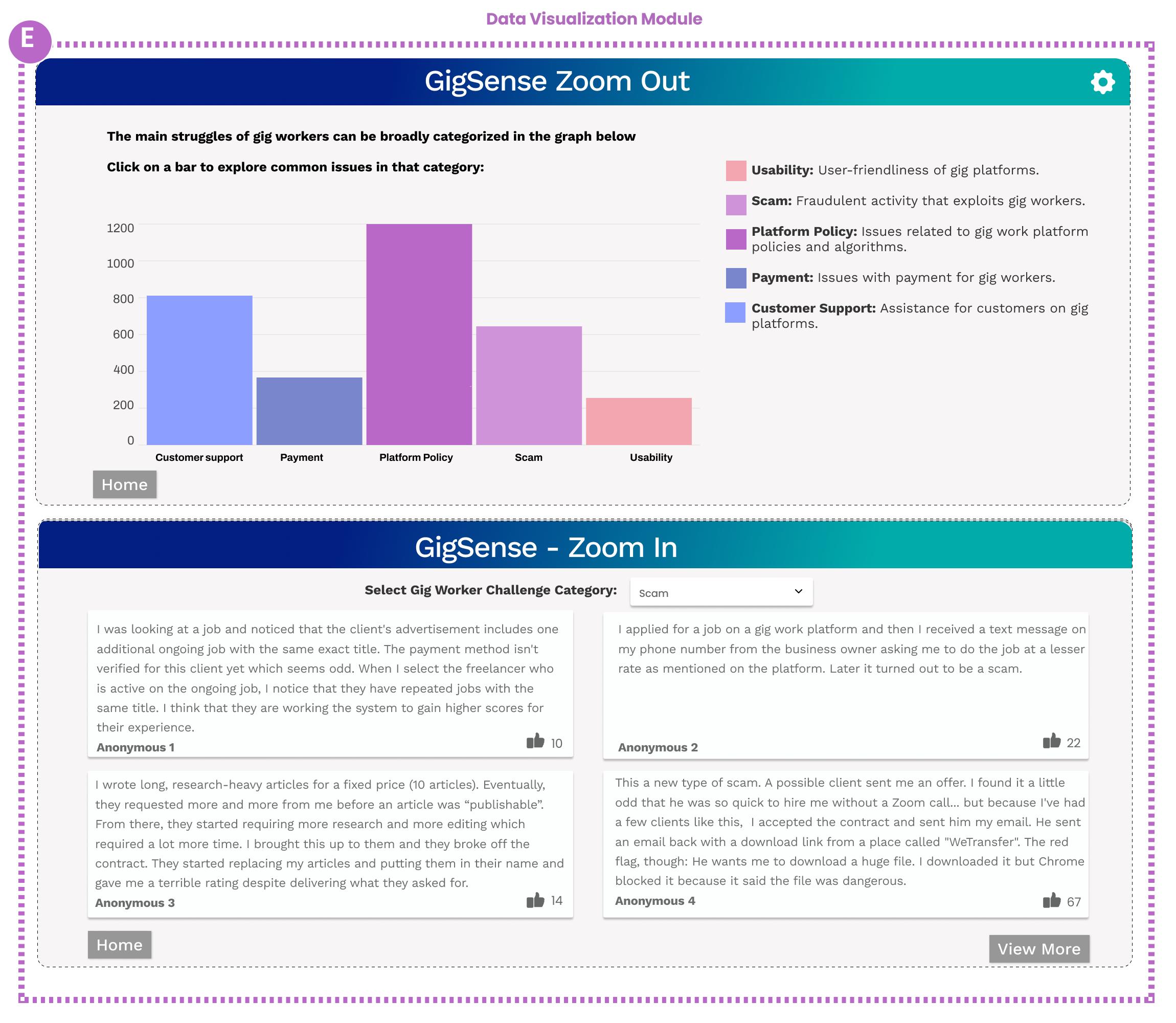}
  \caption{{\small Gigsense's zoom-out and zoom-in view. These views can be accessed by gig workers using the corresponding buttons in the main interface. Gig-sense Zoomed-out view presents an overall view of the problems. Gig-sense Zoom-in view delivers a concise summary of the problem being analyzed, while also enabling users to explore relevant posts associated with that specific challenge.}}
  \label{fig:zoom}
  
\end{figure}

\subsubsection*{\bf Data Gathering Module. } Gig workers supply Gigsense with a roster of subreddits from which they intend to pinpoint potential problems and datasets containing assessments for gig work platforms (This is the ``Step: Search and Filter'' in Pirolli et al.'s sensemaking loop \cite{sentido}). Next, Gigsense connects to the Reddit API to read and extract all the posts from the subreddits that gig workers initially provide. GigSense additionally uses a web scrapper to extract data from reviews left on Apple’s and Google’s app stores by gig workers. Note that our data gathering module only collects reviews that have between one and three-star ratings. The module considers that these review data would represent complaints and problems that gig workers are experiencing. Gigsense also lets workers manually enter issues into the system if they choose to do so (``Step: Read and Extract'' in the sensemaking loop). Using the real-world gig workers' complaint datasets (actual gig workers’ subreddits and complaints)  in our system design aims to bring inclusiveness about gig worker concerns and complaints. The show button allows workers to see a list of all the data sources used in GigSense.

\subsubsection*{\bf Problem Summary Module.} Acknowledging the potential enormity of the data gathered through the Data Gathering Module and its potential complexity for human interpretation, this module centers its efforts on summarizing the data (``Step: Schematize’’ in the sensemaking loop).  For this purpose, Gigsense uses LLMs to facilitate the exploration and sensemaking of the extensive datasets of complaints, enabling workers to categorize and summarize the large datasets into problem categories. This module also has two buttons to allow workers to navigate to the Data Visualization module, which offers a ``zoom-in'' view for meticulous examination of the data or a ``zoom-out'' view for a panoramic grasp of their issues (See Fig \ref{fig:gigsense}.C). 

\subsubsection*{\bf Data Viz Module}  This module enables workers to analyze problems at various levels, aiding their sensemaking process. It utilizes semantic zooming, a method for visualizing data across different abstraction levels \cite{bederson1994pad,suh2023structured,suh2023sensecape}. Workers have access to two views (See Fig.\ref{fig:zoom}) showcasing varying detail levels for problem analysis.
The zoomed-out view of this module presents an interactive chart of the problems (``Step: Schematize'' in the sensemaking loop) where workers can grasp a high-level understanding of the specific types of problems (e.g. Payment, Platform Policy, Scam, Customer Support, Usability, etc) faced by them. Gigsense utilizes LLMs to streamline the sensemaking process by categorizing dataset posts into distinct categories. These categorized posts are later used by the system to create the different graphs and the corresponding category descriptions. The use of LLMs speeds up data organization, allowing workers to engage in sensemaking more efficiently. The interactive chart also serves as a powerful tool for raising awareness among gig workers about their shared challenges. For instance, workers can engage with these graphs by hovering over them, enabling access to the exact number posts (complaints) within each category. This functionality contributes to the worker's sensemaking process by facilitating the easy identification of prevalent issues commonly cited in complaints. 

The volume of information available to workers shifts according to the zoom level. To let the worker delve deeper into individual problems, Gigsense offers a zoom-in perspective. Choosing the zoom-in view shifts the display from graphs to the full text of posts relevant to the problem being analyzed. This view allows workers to view the nuanced aspects of issues that pertain to the topic analyzed and upvote on individual problems. By visually displaying the prevalence of issues (number of upvotes), it creates a sense of solidarity \cite{savage2020solidarity}. Workers can see that they are not alone in their struggles, fostering a collective identity and shared purpose \cite{lee2015more}. It provides a static depiction of the number of workers grappling with problems across different gig work platforms, acting as a catalyst in building communities of gig workers facing similar challenges to encourage collective intelligence. 
The zoomable interface facilitates sensemaking by allowing users to toggle between high-level context and detailed analysis. Note that as \cite{mulgan2018big} discusses, collective intelligence requires focused attention on what matters. The system is thus designed to aid workers in focusing on their problems, allowing them to analyze them from both detailed and broader perspectives.

\subsubsection*{\bf Collaborative Solution Module.} This module further facilitates the sensemaking process and focuses on helping gig workers to develop concrete solutions to address the problem analyzed (``Step: Build Case' '' in the sensemaking loop). It incorporates sub-modules such as the: ``Sensemaking Chat'', ``Shared Document'' and  ``Collaborative Solution Space''. The Sensemaking Chat submodule allows workers to engage in conversations to discuss and investigate the problems they encounter in their work. They can communicate through asynchronous text messages to accommodate different schedules. The  ``Shared Document'' enables workers to understand the existing problems and collaboratively create action plans (solutions) to address them. Gigsense also includes a functionality that allows users to make annotations in the shared document. This way, workers can collectively review and approve their proposed solutions. Finally, the ``Collaborative Solution Space'', just like the sensemaking process, features a space where workers can showcase the final solution they mutually agreed upon \cite{weick1995sensemaking,dervin1983overview}  (``Step: Tell Story'' in the sense-making loop).Note that this module supports collaborative work that initiates early stages of collective intelligence\cite{shaw2014computer}.

\subsubsection*{\bf AI-Enhanced Solution Module.} It is important to acknowledge that for some workers it may be hard to propose solutions \cite{snook2006recruiting,packham2008active}. To address this, the module leverages LLMs to offer workers suggestions on potential solutions and concrete collaboration plans, providing inspiration and initial guidance. However, given our values of prioritizing human connections among workers, AI-generated solutions are presented with lower priority in GigSense's interface. GigSense emphasizes that the AI-generated solution is not the definitive solution and explicitly states that this solution was generated using Generative AI.  Note that in our system, AI-generated solutions are not personalized for individual users. However, the prompts sent to the LLM include all data related to the workers' specific problem. Using real-world gig worker data likely enhances our system's ability to provide contextually relevant AI solutions. Our appendix provides information about all the prompts we utilized for the LLM. Note also that GigSense uses pre-defined prompts to help gig workers effectively interact with LLMs, without requiring specialized skills \cite{Challangestoprompt}. The pre-defined prompts are also used because GigSense is not designed to address all queries, but instead, aims to offer concrete solutions for problem-solving inspiration. Furthermore, workers are informed about the possibility of errors in AI-generated solutions via a disclaimer. This helps not only to ensure that solutions generated by humans receive priority, but also foster more transparent and responsible use of Generative AI while still utilizing AI to assist and inspire workers in their problem-solving endeavors. This module is designed to enhance the ``Step: Build Case'' and ``Step: Tell Story" stages of the sensemaking loop. It achieves this by promoting a collaborative process between humans and AI, fostering co-creation and mutual contribution to the development of solutions.
\\

\section{\bf GIG SENSE: EVALUATION}
The evaluation of GigSense aims to address key research questions:
1) \emph{Speed:} Can GigSense facilitate rapid sensemaking, allowing gig workers to seamlessly contribute to the collective intelligence? 2)\emph{ Contribution:} Can GigSense amplify workers' contributions in sensemaking by enhancing problem identification and solution generation? 3) \emph{Usability:} Does GigSense's AI-enhanced interactive interface bring better user experiences?

\begin{figure}[ht]
  \centering
  \includegraphics[width=0.45\textwidth]{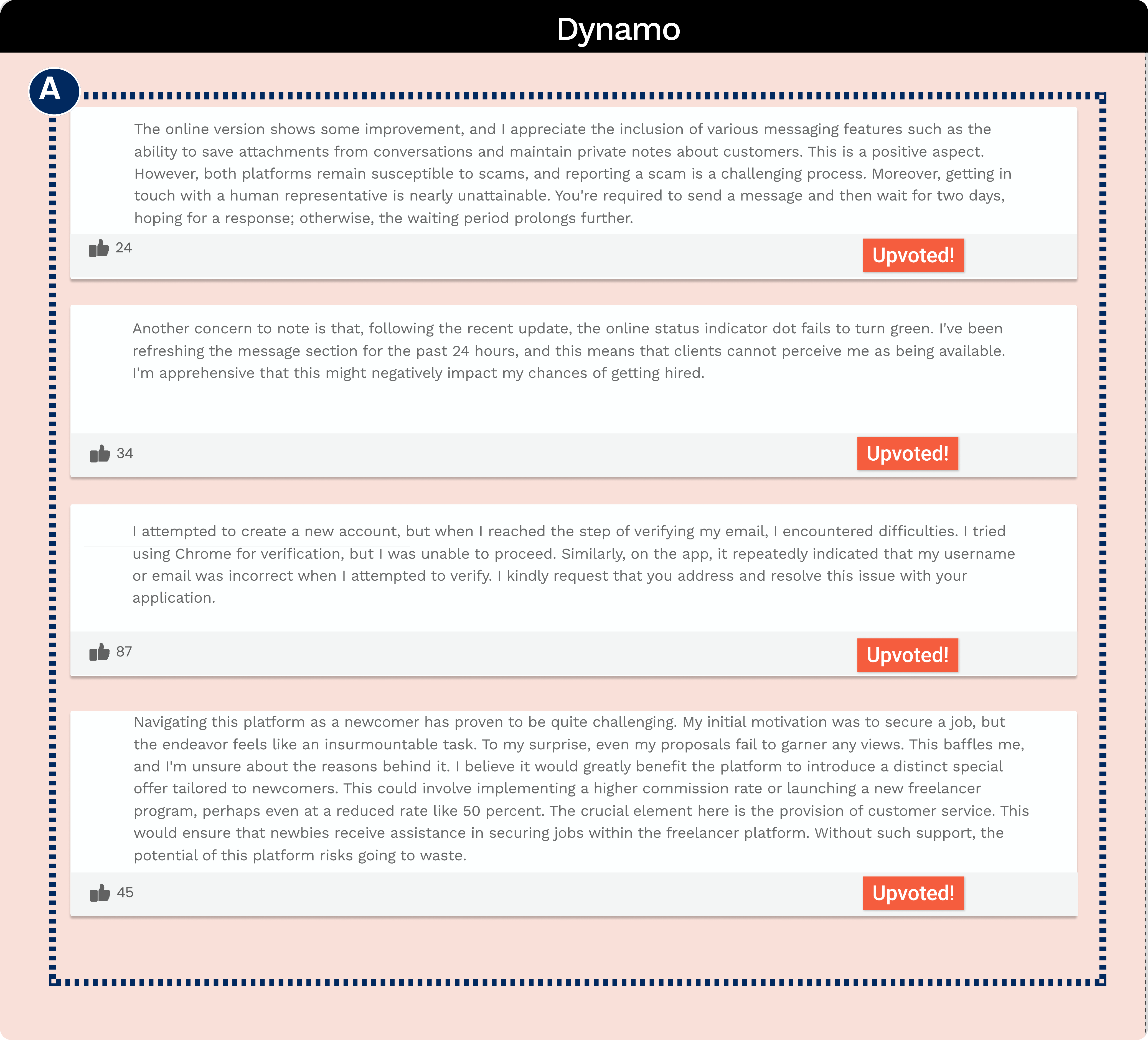}
  \caption{{\small Overview of the control interface.}}
  \label{fig:dynamo}
\end{figure}

{\bf Procedure.}
To study the above questions, we conducted an IRB-approved between-subject user study with 24 participants. We divided the participants into intervention (GigSense condition) and control condition. Participants in both groups were asked to complete the same tasks linked to collective intelligence: pinpointing collective issues and suggesting solutions. Participants in the Gigsense condition used our Gigsense platform to complete the tasks (see Fig \ref{fig:gigsense} and \ref{fig:zoom}, while participants in the control condition used an interface resembling gig platform community forums and features resembling ``We Are Dynamo'' interface \cite{upworkCommunity2024,salehi2015we}. We built ``We Are Dynamo'' to simulate the general functionality of the original system which is no longer available for use. In the control interface, users can engage in the features they would in the original version of 'We Are Dynamo' (such as posting ideas for action and upvoting others' ideas), see Fig. \ref{fig:dynamo}.

Next, we compared the quality of solutions that were generated using GigSense and the control interface. We also studied the usability of GigSense in comparison to the control interface. It is also essential to recognize that human-AI interfaces do not automatically surpass traditional list-based ones in effectiveness. Therefore, it is not clear that the control interface will lead to inferior outcomes. The perceived superiority of AI-enhanced interfaces, often attributed to their informative nature, does not always translate to practical advantage. In fact, interfaces with less information can be preferable, as they help avoid cognitive overload and reduce complexity \cite{appleyard2004paradox,krug2000don,sweller1994cognitive}. Their simplicity, coupled with lower training requirements and ease of use, can make list-based interfaces particularly beneficial in dynamic, time-sensitive work environments \cite{tidwell2010designing,jesse2011elements}. Therefore, in certain scenarios, opting for a streamlined and less informative interface can be a more practical and efficient choice \cite{maeda2006laws,colborne2017simple}. Acknowledging the uncertainty surrounding the most effective design for this scenario, we initiated our user study.

Note that both systems (GigSense and the control condition) used identical datasets encompassing gig worker problems, which were taken from social media posts (subreddits) and reviews on the Google and Apple app stores.  GigSense received the data and leveraged its backend with LLMs and interactive interfaces to offer gig workers a multi-level analysis of the problem space. Similarly, the control interface organized problems based on upvote count, akin to Dynamo's original design where workers can post and upvote short ideas for action. If the idea gets enough upvotes, it turns into a campaign. 

 We arranged the study under the assumption of gig workers operating asynchronously in their collective efforts. This asynchronous setup is crucial due to the varied schedules of gig workers \cite{lehdonvirta2018flexibility}, which might hinder synchronous collaboration. Our aim was thus to ensure effective asynchronous utilization of our tool for seamless completion of collective intelligence tasks. Participants in both conditions engaged with their respective assigned systems and fulfilled the following tasks, drawn from existing literature concerning activities associated with the initial phases of collective intelligence \cite{zhang2014wedo,shaw2014computer,cross2004more,benkler2017peer}.
\begin{enumerate}
\item Provide a summary of one specific problem encountered by gig workers.
  \item Provide a summary of three different problems faced by gig workers.
  \item Enumerate three problems that demand attention due to the adverse impact on workers.
   \item Propose solutions to the three problems you identified that were crucial to be addressed.
  \item Propose a solution to a problem raised by another gig worker.
  \item Propose three solutions to problems raised by other gig workers.
\end{enumerate}

{\bf Participant Recruitment.}
To recruit participants, we generated a job listing on Upwork, extending an invitation to gig workers to join our study. Our selection criteria for participation in the study were workers who:
(a) were aged 18 or above; (b) possessed at least one year of gig work experience (to ensure familiarity with the challenges faced by workers); and
(c) demonstrated proficiency in spoken, written, and comprehended English (to facilitate effective communication with participants). From this, we recruited 24 participants (8 females, 16 males, Median age=27, SD=7.186). After recruitment, we randomly assigned participants to the control and GigSense conditions using the block randomization technique \cite{BlockRandom}. In the end, 12 participants were assigned to the control condition, (P1-P12), and 12  were assigned to the GigSense condition (P13-P24). Participants in our user study were compensated \$10/hr for their participation.

\begin{table}[]
\footnotesize 
\setlength{\tabcolsep}{0.1cm} 
\begin{tabular}{p{1cm} p{1cm} p{1cm} p{2cm} p{1.5cm}  p{6cm}} 
ID & Age & Gender & Race & Work Exp.  & Area of Expertise \\ \midrule
P1 & 48 & Male & White & 4 Years & Content writer, Graphic designer \\

P2 & 30 & Female & Asian & 2 Years & HR, Data Entry, Survey \\

P3 & 19 & Male & Black & 1 Years+  & Book Editing, Copywriting \\

P4 & 32 & Male & Black & 6 Years & Web Developer, Digital Marketing \\
P5 & 38 & Male & White & 7 Years  & Career Coaching, Content Writing, SEO, HR \\
P6 & 22 & Female & South Asian & 2 Years  & Content Writer \\
P7 & 24 & Male & South Asian& 3 Years & Content Writing, Online Education \\
P8 & 35 & Male & White & 7 Years  & Scriptwriter, Book Editing, Content Writing \\
P9 & 31 & Male & South Asian & 4 Years & Software, IT, Graphic Design \\
P10 & 26 & Female & Black & 5 Years & Virtual Assistant, Digital Marketing \\
P11 & 28 & Female & Black & 3 Years  &  Data labelling, Content Writing \\
P12 & 36 & Female & Asian & 4 Years  & Content Writing, Digital Marketing \\
P13 & 27 & Female & White & 4 Years  & Book Editing, Copywriting, Content Writing \\
P14 & 25 & Male & Black & 6 Years  & Human Resource, Career Coaching  \\
P15 & 23 & Male & South Asian & 5 Years  & Software, IT \\
P16 & 40 & Male & Black & 10 Years  & Academic Research, Audio Production \\
P17 & 19 & Male & White & 2 years & Software \\
P18 & 22 & Male & South Asian & 3 years & Digital Marketing, Software \\ 
P19 & 22 & Female & South Asian & 2 years & Content Writing, Graphic Design \\
P20 & 27 & Female & Black & 5 years & Book Editing, Content Writing \\
P21 & 25 & Male & South Asian & 2 years & Software, Digital Marketing \\
P22 & 22 & Male & South Asian & 3 years & Online Education \\ 
P23 & 28 & Male & Asian & 6 years & Human Resource, Digital Marketing \\
P24 & 34 & Male & White & 8 years &Software, Career Coaching \\
\\
\end{tabular}

\caption [Table showing participant demographics in the study including categories like age, gender, race, work experience, and area of expertise of all the participants in the study. First, 12 participants (P1-P12) were assigned to the control condition, and the remaining 12 (P13-P24) were assigned to the GigSense condition.] {Participant Demographics in our study.}
\label{tab:participants}
\end{table}

\subsection*{ \bf Measures and Data Analysis}
We adopted a mixed-method approach, enabling us to harness the strengths of both quantitative and qualitative techniques. Alongside the sociodemographic data, we collected a range of quantitative metrics to address our three research questions related to speed, contribution, and usability. To provide a deeper perspective and also contrast against the control condition, we complemented our quantitative findings by conducting exit interviews with participants in both conditions. Through this approach, we gained invaluable insights into their user experiences.

\subsubsection*{\bf Metric: Speed} In both the control and GigSense conditions, participants used a button to signal task start and completion. The systems recorded timestamps for each button press, enabling precise tracking of task durations per participant.

\subsubsection*{\bf Metric: Contribution} 
To determine if GigSense improves workers' contributions to sensemaking and collective intelligence, we evaluated its impact on helping workers identify problems and develop various and feasible solutions. Studying these metrics aligns with prior research on investigating the effectiveness of different sensemaking interfaces \cite{siangliulue2016ideahound}, and identifying problems and proposing solutions \cite{zhang2014wedo,shaw2014computer}.

For feasibility evaluation, we gathered all problems and solutions from each condition and hired three English-speaking, college-educated raters through Upwork. They independently evaluated each proposed solution's feasibility on a 7-point Likert scale, considering the problem and solution. Feasibility was assessed based on the viability of the proposed solution in effectively addressing the issues encountered by participants. Inter-rater reliability for each group was measured using a two-way mixed Intraclass Correlation Coefficient (ICC), revealing excellent agreement among raters: 0.93 for GigSense and 0.94 for the Control group. Overall, this analysis effectively demonstrated the impact of our interface on workers' sensemaking, as assessed through their contributions to identified problems and solutions.

\subsubsection*{\bf Metric: Usability.} To assess participants' views on GigSense's usability and compare it with the control, we employed the System Usability Score (SUS) \cite{brooke1996sus}, a validated metric. The SUS is comprised of 10 questions on a five-point Likert scale. It is widely used for measuring usability and comparing systems \cite{peres2013validation,lewis2018system}. It offers valuable insights into users' subjective experiences with a given technology. Following participants' interaction with their assigned system (control or GigSense), they received the SUS questionnaires. We then computed the SUS scores reported by participants for their respective systems. Following this, each participant provided a usability score for the system they used in their assigned condition. 

\subsubsection*{\bf Qualitative Study.}
To augment our quantitative data, we conducted exit interviews for richer insights into participants' user experience, adding a qualitative dimension. These interviews captured participants' feedback and thoughts about the respective interfaces they used. Interview data was transcribed and subjected to open coding \cite{mihas2019qualitative}. This process involved deriving initial concepts from the data through a mix of bottom-up and top-down theme extraction. Two of the paper's authors independently conducted bottom-up coding, resulting in 13 axial codes, which were then applied top-down to all interview transcripts.

\section{\bf Results}
Next, we proceed to unveil both the quantitative and qualitative results that surfaced during our evaluation.

\subsection{\bf Quantitative Results.}
\subsubsection{\bf Time}
\begin{figure}
    \centering
    \includegraphics [width=\textwidth, alt =Four charts showing study results. a) Median time to complete tasks was lower for GigSense users. b) GigSense users proposed more feasible solutions. c) GigSense users were faster at tasks. d) GigSense had higher system usability scores.] {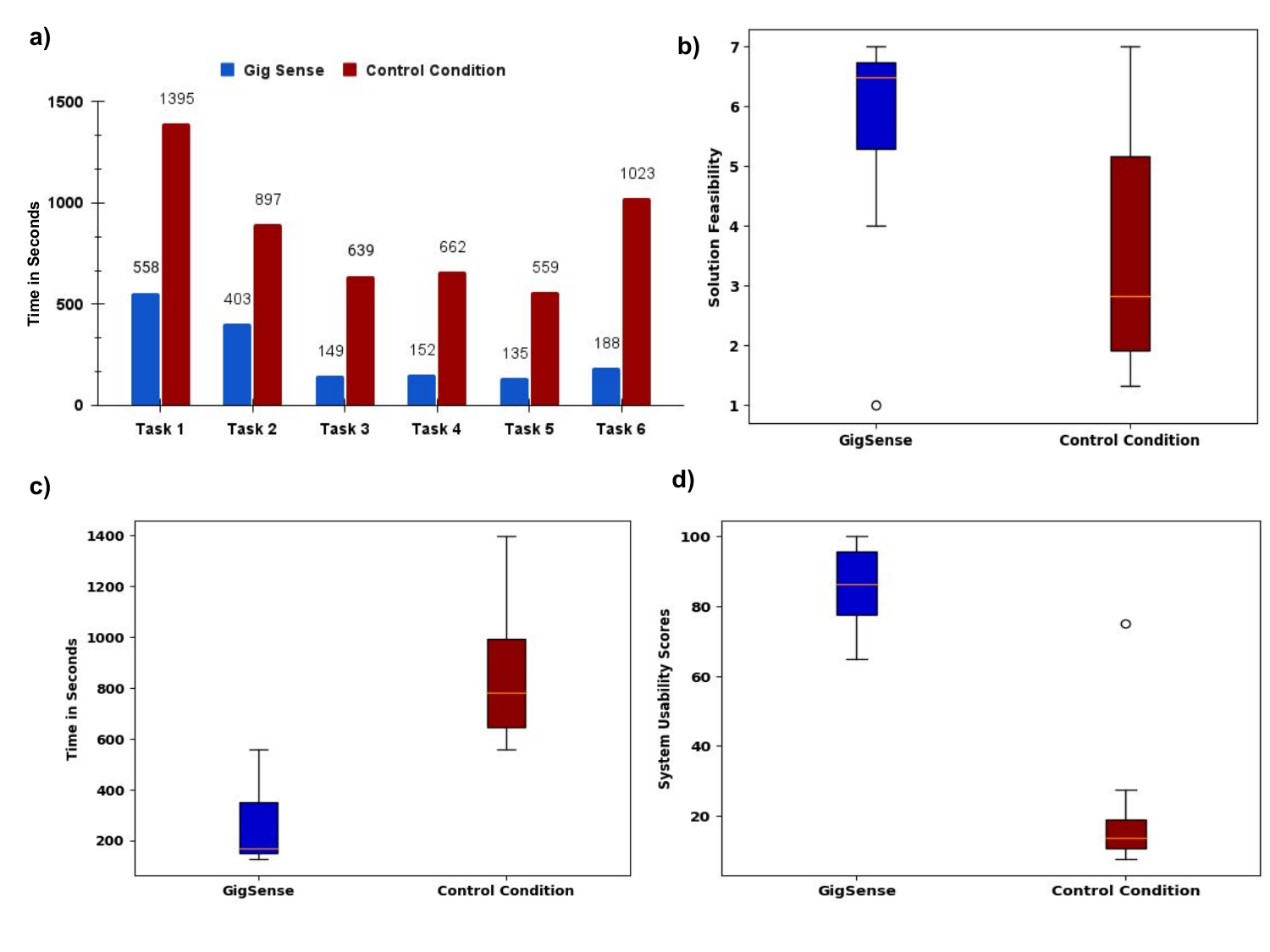}
    \caption{Overview of: a) Median amount of time gig workers in each condition took to complete the different tasks; b)Box plot showing the evaluation of solutions proposed by gig workers in both groups based on their feasibility (on a 7-point Likert scale);
    c) Box plot showing task completion time in both groups; d) Box plot showing System Usability Scores in both groups}
    \label{fig:timeSUS}
\end{figure}

Time can pose a challenge for gig workers aiming to engage in collective intelligence, as not all workers enjoy the luxury of allocating extensive time to this pursuit \cite{packham2008active}. To address this concern, we assessed the duration participants required to accomplish the different problem-solving tasks defined in our study. Figure \ref{fig:timeSUS}.a) provides a comprehensive depiction of the median time taken by participants to complete the entire set of tasks in both conditions. (Figure \ref{fig:timeSUS}.c) depicts the box plot for both groups.
 The results of our study indicate that participants in the GigSense group exhibited faster task completion times  (Mean 264.08 seconds, Median=170 seconds, SD= 175.45 seconds) compared to the control condition (Mean= 862.5 seconds, Median=779 seconds, SD= 313.93 seconds). To study whether these differences between the GigSense condition and control were significant, we conducted appropriate statistical tests. First, since our data did not meet the assumption of normality we employed the Mann-Whitney U test, a non-parametric test specifically designed to compare the medians of task completion times between the intervention group (GigSense) and the control group. The Mann-Whitney U test revealed a statistically significant difference between the two groups in our study, with a p-value of 0.002. This p-value indicates that there is a statistically significant difference between the two groups in our study. This implies that gig workers were significantly quicker in their problem-solving tasks when utilizing GigSense compared to traditional interfaces. Overall, the data provides evidence that GigSense offers a promising approach (RQ1) to improve task completion times in problem-solving tasks related to collective intelligence.

 \subsubsection{\bf Contribution. (Evaluating Gig Workers' Solutions)}
To evaluate GigSense's effectiveness in enhancing workers' contributions to sensemaking and collective intelligence, for both groups, we assessed the number of identified problems, the number of proposed solutions, and their feasibility.

First, we observed that the GigSense group identified more problems compared to the control group. Specifically, the GigSense group had a median of 10 problems identified (mean 9.58) versus 6 (mean 6.33) for the control. A Mann-Whitney U test revealed this was a statistically significant difference (p = .01047). Similarly, in assessing the solutions proposed, the GigSense group suggested a higher number of solutions. In particular, they had proposed a median of 10 solutions (mean 9.41) compared to 6 (mean 6.33) for control - a statistically significant difference (p = .01379).

To study whether GigSense effectively supports the generation of more feasible solutions for gig workers, we conducted an expert evaluation of the solutions that participants in both groups proposed.  We found that gig workers in the GigSense group produced in general more feasible solutions (Median=7 [``Very Feasible''], Mean=5.76 [somewhat feasible], SD=1.8) than workers using the control interface (Median=3 [Slightly Unfeasible], Mean=3.58 [Slightly Unfeasible], SD=2.1). We plotted a box plot graph (\ref{fig:timeSUS}.b) to better visualize the differences in the solutions each group contributed. Next, we wanted to identify whether the differences in the feasibility of solutions were significant. 
Through our analysis, we first identified that the distribution of feasibility scores did not meet the assumption of normality. Consequently, we again performed the Mann-Whitney U test. The results of this test indicated a statistically significant difference between the two groups, with a p-value of 0.248. This suggests the presence of a significant distinction between the feasibility of the solutions that gig workers contributed in the GigSense condition and the control condition.

In conclusion, our findings reveal that GigSense facilitates the contribution of more feasible solutions (RQ2) by gig workers, as evidenced by the significant difference in the expert evaluation scores between the two groups. GigSense also enhances sensemaking capabilities, given the GigSense group identified more problems, and proposed more feasible and plentiful solutions, compared to the control group.

\subsubsection{\bf System Usability Scale}
Utilizing the System Usability Scale (SUS) \cite{brooke1996sus,empiricalSUS,peres2013validation}, we studied the reported usability levels of GigSense among gig workers and drew a comparison with those reported for the control condition. Figure \ref{fig:timeSUS}.d presents the boxplots for the System Usability Scale scores of GigSense and the control condition. Our findings revealed a notable trend: the median SUS score for GigSense (Mean=86.25, Median=86, (adjectival rating: Excellent), SD=11.6) was higher than the median SUS score for the control condition (Mean=20.41, Median=14, (adjectival rating: Poor), SD=18.7). Building upon this observation, our subsequent focus was to determine the significance of this disparity. As the SUS scores did not meet the assumption of normality. We therefore opted for the Mann-Whitney U test once again to compare the medians of SUS scores between the GigSense condition and the control condition. The analysis revealed a statistically significant difference between the two groups in our study(RQ3), with a p-value of 0.001.

\subsection{\bf Qualitative Results}
To analyze, the interview response, the transcripts were open-coded individually by two researchers and further axial coding was performed. From 13 top-down axial codes, we developed four themes that organized the main insights from our exit interviews. We labeled participants' responses within the "GigSense group" as \textit{"PGS"} and those in the control group as \textit{"PC"}.
\subsubsection {\textbf{Collective Intelligence via sensemaking}}
GigSense assisted workers in recognizing shared problems and devising solutions. It assisted workers in initiating collective intelligence via sensemaking processes. Especially as the interface offered by GigSense facilitated the exploration of problems from various perspectives,  a key step in the journey towards collective intelligence). This, in turn, led to an awareness of the severity of specific problems:
\textit{``I think it [GigSense] can bring people a sense of unity and frustration. So like, it's nice to be able to see, especially all these [pointing at GigSense's graphs showcasing the magnitude of problems faced by workers], because you can see how bad it [a specific problem] is. Then here, [the worker clicked a problem and zoomed into the specific things other workers complained about the problem], all of these reviews [workers' complaints about a given problem], show this problem is super common [...] the reviews  [workers' complaints about a given problem] are like written in honest frustration. And it seems kind of like a recognition of: ``Oh, everybody's really got this problem'' and frustrated by these things....}  (PGS 8). GigSense showed workers their challenges weren't just experienced by themselves. This realization likely spurred action: \textit{ “There could be a time like, you might be facing a problem. But you might be thinking that that's something that you are only facing. And it could be a problem related to usability or payment. You do nothing about it. But when I see this thing on your website [GigSense], you see, other people are also facing it. So this is not just you since there are others like you [facing similar problems]. You get to talk to each other through your shared concerns"} (PGS 10). Recognizing shared challenges likely reinforced workers' collective identity, potentially inspiring them to undertake actions that benefited them as a whole.

\subsubsection{\textbf{Assisting Workers in Solution Generation.}} Participants appreciated how GigSense effortlessly enabled solution generation. They particularly liked the fusion of the AI-enhanced interface with the interactive collaborative space, as it seamlessly facilitated the combination of existing solutions into more fortified ones: \textit{``I like this part [pointing to GigSense's AI-enhanced solution space and the button for generating new solutions]. I love that I can interchange the suggestions that I want anytime [Here, they highlighted the collaborative solution space]”} (PGS 5). 

Participants also valued how GigSense empowered them to formulate solutions for less familiar problems, broadening their capacity to participate in problem-solving across areas where they might not usually contribute ideas: \textit{“Often I see a problem that I am not familiar with. The AI suggestion kind of suggests you somewhat relatable thing, so in a situation where you might not have anything to contribute, you still have something to suggest, I think that's really cool.”} (PGS 11).

Given the diverse backgrounds and regions of gig workers \cite{berg2015income,berg2021working}, AI assistance played a crucial role in helping participants overcome language barriers, facilitating the generation and seamless sharing of solutions among them: \textit{ ``The AI is very useful for me because it gives me the many ideas needed, and I also struggle with English, and many times I can't put clearly what I am trying to say''} (PGS 6).

\subsubsection {\textbf{Collaborative Problem-Solving}}
Our system's interface was well-received for its ability to help people come together and drive collaborative problem-solving efforts among gig workers: 
\textit{“forums [traditional tools] can just be people complaining at each other and nothing happening, whereas this [GigSense] kind of gathers data and proposes solutions. It is just more of a solution based, I think it helps visualize problems [...] then the solutions can be given or can be received by somebody that says: ``Oh, let me make something [a solution]'' [...] It seems like a space for the solutions to kind of come together”} (PGS 8). Participants also underscored the potential of GigSense to empower them in joining forces, collaboratively suggesting solutions that encompass technological enhancements for gig platforms and necessary policy changes. Participants believed that these combined efforts could cultivate a more nurturing environment for gig workers: \textit{“So, with these solutions [the ones generated collectively through GigSense] either that is, somebody creates a new and better platform or one of the platforms reads it [solutions generated collectively through GigSense] and updates their policies and changes things to make it more friendly for the Freelancers. I think it's like it could be space for creating better paths for both freelancers and platforms”} (PGS 12). Participants also expressed that our tool facilitated stronger connections among fellow workers, fostering a sense of empathy for shared challenges. This, in turn, with the help of the interactive interface, bolstered their collaborative problem-solving efforts:
\textit{"I resonate with what is being said [workers sharing their problems] as somebody that's been doing gig work for a while. I highly resonate with it [...] And this interface makes me able to put in my words for [creating] the solution"} (PGS 8).

\subsubsection {\textbf{AI, Humans, and Interface Design.}}
We crafted GigSense's interface to prioritize the placement of human-generated inputs—solutions proposed by gig workers—above AI-generated suggestions. Participants mentioned that they appreciated this design, highlighting that they liked seeing solutions from both human and AI sources, but especially appreciated the emphasis on those contributed by fellow humans: \textit{“It is good to get the ideas [solutions] from the AI. But I liked that some live answers were there [solutions proposed by gig workers], and some live people, real human beings answered it [gig workers provided solutions]”} (PGS 4).

On the other hand, LLMs occasionally produce content with errors. Our concern revolved around gig workers unquestioningly adopting the solutions provided by the LLM. In our interface design, we proactively introduced an additional measure by including a disclaimer regarding AI-generated suggestions. This strategic move aimed to encourage workers to approach the AI's recommendations thoughtfully, rather than hastily adopting them. Our primary objective was to avert any unforeseen repercussions stemming from thoughtless adherence to AI advice. Participants valued the implementation of this design approach, which underscores our commitment to responsible AI utilization: \textit{“The solutions, (AI suggestions)  can be helpful to guide you, but cannot provide you with a perfect solution. That's already there in the disclaimer, So you can get ideas. But you still have to use your brain and experience to answer”} (PGS 2). Participants in general expressed that GigSense's interface gave them a sense of autonomy in their decision-making, as they were not constrained to unquestioningly adhering to the AI-generated solutions:
 \textit{“I like the fact that the system is suggesting a solution, not completely telling me this is exactly the solution for this problem”} (PGS 5). Participants also conveyed a sense of resonance with GigSense's AI-generated solutions, emphasizing that they were not out of place within GigSense's interface: 
\textit{“The AI suggestions continuously synced with me, while I'm thinking of my things to write. The answers all seem aligned with my thoughts and were really helpful suggestions" } (PGS 3).

Workers also valued how GigSense's intelligent interface empowered them to methodically structure and analyze problems in diverse ways: \textit{“..the bar chart [bar chart that was automatically generated by GigSense to show number of worker messages generated about a particular problem] that like immediately told me what the stats were, and then it was quite straightforward to go through each complaint [workers' messages about the problem] and understand all that stuff behind the problem”} (PGS 4). This type of interface capability was absent in the control condition and participants complained about it: \textit{“I keep on scrolling, and there are different problems. I am not able to categorize it [categorize worker messages about a particular problem]. So it is time-consuming. I have to scroll through each and every message”} (PC 9). Similarly, some of the workers on the control group imagined better futures where they could have access to some of the interface features of GigSense, such as problem categorization:  \textit{“I think it would have been better if the categorization of problems was there [on the control interface]. It breaks down things easier"}(PC7). Other workers in control also desired some of GigSense's interface features, particularly those about concise summaries and the ability to conduct sensemaking in shorter periods of time: \textit{“So it will be very difficult to read those 1,000 words line by line. So it's specifically if you use some artificial intelligence or something, the whole thing can be summarized in 200 words. The lesser it is the the less time it will consume, and the user can easily understand.”} (PC9). In contrast, participants in the GigSense condition highlighted how they enjoyed that GigSense's interface helped them to minimize sensemaking costs: \textit{“I can read the issues that gig workers are facing from here [The worker pointed to the summary module of GigSense], and I don't have to open them [worker messages about a particular problem] one by one. I think it's helpful. So, since gig workers are very busy, having a summary here is very helpful [...] I can read all of this in under a minute} (PGS2).
\section{\bf Discussion}
Our user study showcased that GigSense users generated solutions for collective issues significantly faster, with a significant increase in perceived usability, and a significant enhancement in the feasibility of these solutions. These outcomes provide valuable insights into the role of LLMs in supporting sensemaking processes within collective intelligence. Here, we discuss ongoing challenges and prospects for sensemaking tools for problem-solving with gig workers.

\subsubsection*{\bf Powering  Collective Intelligence}
In pursuit of enhanced problem-solving, GigSense strategically incorporated the advanced capabilities of Large Language Models (LLMs) to streamline and optimize the sensemaking process. For instance, participants valued how GigSense empowered them to formulate solutions for less familiar problems, broadening their capacity to participate in problem-solving across areas where they might not usually contribute ideas. This correlates with earlier studies that have indicated LLMs' ability to generate notably superior ideas compared to humans \cite{girotra2023ideas}. GigSense, aided by LLMs, allowed workers to explore and generate solutions for less familiar problems, broadening the scope of solutions generated, analogous to the process of searching for relevant information in diverse data sources in the sensemaking phase ({\it ``Step: Search and Filter''}). Moreover, participants appreciated how GigSense effortlessly enabled solution generation. This ease of generating solutions emphasizes the efficiency and user-friendly nature of GigSense's interface, which was further enhanced by the integration of LLMs. Additionally, LLMs contributed to the rapid extraction and presentation of solutions, aligning with the sensemaking step of extracting valuable information, analogous to the ({\it ``Step: Read and Extract''}) in the sensemaking process. In fact, participants expressed that they found it easier to derive solutions swiftly from the information provided within the GigSense platform, facilitated by LLM-driven capabilities. Moreover, Gigsense use of LLMs facilitated the summarization of problems, (“Step: Schematize”) in the sense-making process, resulting in enhanced comprehension and consequently enabling
the proposal of better solutions. For instance, participants using GigSense expressed their satisfaction with how the
platform’s interface facilitated their comprehension of gig workers’ challenges about the topics presented on the
platform. They appreciated the “Problem Summary Module”, which spared them from the tedious process of opening and reading through the long list of individual worker complaints about specific problems. This improvement significantly aided workers in streamlining this task and in making information gathering and summarization much more efficient.  In fact, participants expressed that they found it easier to derive solutions swiftly
from the information provided within the GigSense platform, facilitated by LLM-driven capabilities. These user experiences resonate with prior studies demonstrating how LLMs can enhance users' information analysis capabilities \cite{suh2023structured,suh2023sensecape}   In its current iteration, GigSense did not integrate Large Language Models (LLMs) within the specific steps of  ({\it ``Step: Build Case''}) and ({\it ``Step: Tell Story''}) in the sensemaking process. Instead, it supports these steps via its \textit{Collaborative Solution Module}, allowing users to interact and discuss, and upvote the different solutions generated by them and the AI-suggested solutions. Future work could explore the incorporation of LLMs into these pivotal sensemaking steps to further optimize the process. For instance, integrating LLMs within the ({\it ``Step: Build Case''}) could involve utilizing their language generation capabilities to assist in constructing a comprehensive case or argument based on gathered data. In the ({\it ``Step: Tell Story''}), LLMs might aid in synthesizing and articulating narratives or insights drawn from the information collected, enhancing the storytelling aspect of sensemaking.

\subsubsection*{\bf Catalyzing Inclusive Problem-Solving.}
GigSense is designed to facilitate gig workers' participation in problem-solving. In designing GigSense, we prioritized the unique time constraints faced by gig workers, acknowledging their often limited availability due to potential financial hardships \cite{gray2019ghost,wood2019good,hsieh2023co}. Our aim was thus to ensure quick sensemaking, enabling more rapid production of solutions for collective issues. Balancing this aspiration with the production of feasible solutions presented a challenging task for GigSense. Our user study demonstrated that GigSense indeed yielded more feasible solutions compared to the control interface. A likely contributing factor was that GigSense's interface empowered workers to swiftly assess the zoomed-in and zoomed-out dynamics of their problems. This likely led workers to have a better understanding and thus generate more attainable solutions, compared to list-based interfaces. However, acknowledging that not all workers might prioritize in-depth problem exploration is essential. To address this, incorporating informative messages within GigSense could enlighten users about the benefits of investing slightly more time in analyzing and comprehending problems. 

However, it is also important to highlight that GigSense's design does aim to counter the risk of overlooking nuances in problem-solving by featuring both zoomed-in and zoomed-out interfaces, ensuring a comprehensive analysis. The zoomed-out view provides a comprehensive perspective, crucial for understanding how singular issues are interwoven into broader systemic patterns. This helps workers in recognizing overarching trends and contexts. On the other hand, the zoomed-in interface facilitates a detailed examination of specific problem aspects, allowing for a thorough analysis of individual components. This dual-mode approach effectively balances a macro and micro perspective, ensuring that complex issues are not oversimplified. Consequently, by enabling workers to effortlessly toggle between these views, GigSense enhances their ability to engage in more effective sensemaking and problem-solving. Our user study underscored this, showing that workers using Gig Sense identified significantly more and varied problems, and proposed more diverse and useful solutions, compared to the control group.

GigSense's design also raises a thought-provoking discussion about its potential role in promoting ``technosolutionism," the idea that technology can swiftly resolve complex design issues without deeply engaging with their intricacies \cite{morozov2013save}. This perception can stem from GigSense's design goals to expedite solution-finding, particularly under the constraints of workers who often lack the luxury of time for problem-solving due to their need to focus on livelihood-sustaining activities \cite{gray2019ghost,wood2019good,hsieh2023co}. To start to address this conflict, our GigSense design was inspired by collective intelligence research \cite{mulgan2018big}, highlighting the critical role of interfaces that support focused collaborative problem-solving. Consequently, GigSense's interface was crafted to enable workers to concentrate on specific issues, offering the tools to examine these problems from various deep perspectives. This strategy can hopefully enable workers to delve into their problems  with greater depth and less superficiality.

Based on these ideas, GigSense focuses on orienting workers towards solution-driven actions, actively engaging workers in the process of change, and reinforcing their sense of agency - a core tenet for promoting collective intelligence \cite{mulgan2018big}. Contrastingly, interfaces with a problem-focused approach might induce in workers feelings of helplessness or passivity \cite{seligman1972learned,abramson1978learned}, thus diminishing workers' agency \cite{peterson1983learned,seligman2011building}. This is why we chose not to limit ourselves to a solely problem-focused interface. Nevertheless, GigSense does not disregard the importance of understanding problems in depth. Its interface, designed for both zooming in for detailed problem analysis and zooming out for a broader perspective, supports a more nuanced engagement with problems. This functionality proved effective in our studies, where workers using GigSense identified a wider array and greater number of problems compared to those using the control interface. To summarize, GigSense not only simplifies the journey towards finding solutions, but it also promotes a comprehension of the issues at hand.

\subsubsection*{\bf Collaborative Problem Solving with Human-AI Interaction} Our system introduced a collaborative problem-solving process that integrated human-AI interactions, with a primary objective of enhancing human creativity by leveraging LLMs to empower workers to devise creative solutions to their challenges. This approach complements prior research on LLMs' assistance in enhancing human creativity \cite{cai2023designaid,siangliulue2015toward,kulkarni2013early,hope2022scaling}, emphasizing their supportive role rather than substituting human involvement\cite{cai2023designaid}. GigSense demonstrates the value of designing interfaces that harness the power of LLMs to augment and streamline the sensemaking process to empower non-experts to utilize LLM technology for collective problem-solving. Our results reveal that LLMs supported gig workers (non-experts in the technology) in generating solutions, but our human-AI design ensured workers did not rely solely on the LLM output. Instead, workers used it to complement human-generated content, considering LLM suggestions as one of many sources they could incorporate. For this purpose, we strategically positioned LLM outputs below human-generated content and provided disclaimers about their reliability. Unlike previous studies \cite{harrer2023attention}, our participants welcomed LLM suggestions, incorporating it into their sensemaking process for creating solutions that improved their collective intelligence. Nonetheless, unexpected LLM outcomes could potentially hinder workers' sensemaking and solution production. Future research could explore new human-AI interfaces for addressing problematic LLM outcomes, as well as study interface designs that prioritize different types of solutions based on workers' needs, e.g., novel solutions vs feasible solutions. Notice that the design of the human-AI interactions could influence the nature of generated solutions. Future research should consider recent studies on designing interactive interfaces to explain large language model responses \cite{jiang2023graphologue,huang2023causalmapper}. This transparency can enhance collaboration between end-users and AI-generated solutions. 
\subsubsection{\bf Responsible Use of Large Language Models} 
Employing LLMs demands substantial computational resources, leading to energy-intensive processes \cite{hoffmann2022training,li2023survey,finn2016connection}. It is crucial to assess resources for training and fine-tuning if needed \cite{naser2023we}. However, large pre-trained models like GPT-3 can be highly efficient post-training \cite{liang2022holistic}, and techniques like model distillation can further economize costs \cite{dasgupta2023cost}. It is also important to consider how LLMs may perpetuate biases  from their training data \cite{ferrara2023should}. Recent research underscores intervening to address biases \cite{thakur2023language,dhamala2021bold}. Future research could investigate collective intelligence interfaces designed to enable workers to collaborate in identifying and mitigating biases in LLMs \cite{piktus2023roots}. In connection with this, it is crucial to acknowledge that although GigSense can alert users to errors in LLMs, the broader consequences of disseminating incorrect solutions remain understudied. This underscores the necessity for additional investigations into human-AI collaboration and the implementation of frameworks that prioritize responsibility. \cite{kulkarni2023llms}.

{{\bf Limitations.}}
Our study has several limitations, including a diverse but small sample size of gig workers with varying skills and geographic backgrounds, potentially affecting generalizability. We addressed this by requiring participants to have at least a year of gig work experience, ensuring they could meaningfully evaluate GigSense. Despite its smaller scale, our mixed-methods approach allowed for a rich, qualitative understanding of gig workers' interactions and experiences with the system. Future research could aim to validate our findings across broader samples and investigate the standalone impact of LLMs and interactive interfaces on solution generation for workers. We also integrated real-world gig worker data to broaden the context of challenges faced, acknowledging that not all problems can be solved even with tools like GigSense. Future studies could differentiate between solvable and complex challenges and assess the potential of LLMs and interactive interfaces for complex problem-solving. Although our study offers in-depth insights, a longitudinal study could provide additional data on social network analysis facilitated by GigSense. Upon publication, we will open-source GigSense, inviting further research on AI for collective intelligence. We recognize that LLMs can produce erroneous solutions, and we have aimed to implement safeguards by prioritizing human input and warn of potential errors. However, the introduction of LLMs might suggest a broader solution capability than feasible, indicating the need for future studies to refine user guidance on LLM limitations.

\section{\bf Conclusion}
This paper presents GigSense, an AI-enhanced tool designed to help gig workers collectively understand and tackle their challenges. GigSense enables rapid sensemaking, reducing the barriers to collective intelligence, and fostering effective problem-solving and solution generation among workers. Our study revealed that users of GigSense identified more problems, and proposed solutions that were quicker, more plentiful, and more feasible than those in the control group. Users also reported enhanced usability experiences, valuing GigSense's support in problem understanding, collaborative solution crafting, and its integration of AI with a focus on human-driven solutions. The favorable feedback and functionality of GigSense highlight its capacity to transform gig workers' approach to challenges, turning individual struggles into collective successes.

\begin{acks}
This work was partially supported by the NSF Grant FW-HTF-19541. We also want to thank Swati Agarwal, the Global Partnership in AI, Professor Seth Harris, Jianhua (Chandler) Che, Northeastern's Apprenticeship program, and all the workers who participated in our study.
\end{acks}

\bibliographystyle{ci-format}
\bibliography{Main}

\appendix
\section{\bf Appendix}
We present the prompts we encoded in our system to generate the: (1) categorization of gig workers’ problems; (2)  summaries of their problems;  and (3) solutions generated with OpenAI’s GPT-4 API. 

\subsection{\bf Prompts for categorizing Gig Workers’ Problems}
Our data-gathering module focuses on collecting data about different potential problems faced by gig workers. It has webscrappers to collect data from different gig worker subreddits (r/Freelancers, r/UpWork), as well as the reviews that people left on Apple’s and Google’s app store data about major gig platforms (particularly: Amazon Mechanical Turk, Fiver, UpWork, Taloka, and Freelancer). Note that our data gathering module only collects reviews that have between one and three-star ratings. The module considers that the review data would represent complaints and problems that gig workers were experiencing. Next, our module connects with OpenAI’s API in the backend of our system to first categorize the data into different categories of gig work challenges. To obtain such categorization, We prompted:

“Please categorize the following data into specific problem categories. The goal is to sort the large chunk of data into distinct categories for better comprehension. You should have a problem category with a name and a list of statements related to the problem. If you find a category or theme that is not a problem, do not include it”

\subsection{\bf Prompt for Summarization}

“Given the following dataset of  complaints related to X (where X represents an identified problem category), please provide a concise summary of the problem to facilitate a comprehensive understanding.”

\subsection{\bf Prompt for generating solutions}
The categorized data was then again fed into the backend of our system where we encoded the following prompt:

"Given the following dataset of complaints related to X (where X represents an identified problem category), please provide solutions to each problem category

\subsection{\bf LLM Response} 
 \textbf {Summary of Problem-related to Platform Policy} 
The common theme among the listed 'Platform Policy' problems faced by gig workers is dissatisfaction with the platform's fees and policies. Gig workers are unhappy with the high fees, especially for newbies who cannot afford them, and the arbitrary increases in fees for more experienced freelancers. There are complaints about the cost of connects, commission charges, and withdrawal policies. Additionally, some gig workers are unhappy with the platform's lack of transparency regarding proposal views and analytics. Overall, these problems suggest that the platform prioritizes profits over the well-being of freelancers.\\

\textbf{Solution suggested by LLM}\\
\begin{enumerate}
    \item  Negotiate platform fees: Gig workers can come together and negotiate with the platform for lower fees or a fairer fee structure. They can explain their challenges and how high fees affect their ability to earn a living. With a collective voice, they may be able to persuade the platform to revise its policies.
    \item Seek alternative platforms: Gig workers can explore other platforms that charge lower fees or offer more benefits. They can research and compare platforms to find the one that best suits their needs. Switching platforms may help gig workers find better-paying jobs and reduce the impact of high fees.
    \item Advocate for fair policies: Gig workers can form advocacy groups or join existing ones to lobby for fairer policies for all gig workers. They can work with policymakers to push for regulations that protect gig workers' rights and ensure a level playing field.
    \item Share knowledge and resources: Gig workers can collaborate and share knowledge and resources to improve their chances of finding better-paying jobs. They can form communities and support groups where they share tips, advice, and best practices for finding and landing jobs. By sharing their experiences, gig workers can help each other overcome common challenges.
    \item Encourage transparency - Gig platforms should encourage clients to be transparent in their job postings, payment processes, and communication with gig workers. Clients should disclose all the necessary details related to the job upfront, and payment processes should be clear and easy to understand. This will help gig workers to identify any red flags and avoid working with fraudulent clients.
\end{enumerate}

\textbf{Summary of Problem related Usability} \\
This category of problems is related to the usability of gig work platforms. Gig workers are facing issues with the web version and mobile apps. Some workers are complaining about missing features or outdated versions of the app, leading to delays and incorrect data. Others report problems logging in, receiving error messages, and being unable to download or install the app. Several users have encountered issues with the messaging system, including delays in loading messages, formatting issues, and glitches that make it difficult to send and receive messages. Many workers have also reported difficulty applying for jobs, with some issues including the page refreshing after attaching files, lost text inputs, and difficulty retracting proposals. Additionally, there are reports of notifications being delayed or not working, as well as issues with the app crashing or freezing. \\

\textbf{Solution suggested by LLM} \\

\begin{enumerate}

 \item Collaborate with other gig workers: Gig workers can form groups or communities where they share information and experiences about different gig platforms. By doing this, they can help each other find the best-paying gigs and avoid low-paying ones. This collaboration can also help them negotiate better rates and resolve conflicts with clients.
\item Leave feedback and reviews: Gig workers can leave honest feedback and reviews about the gig platforms they have used. This can help other gig workers make informed decisions when choosing which platform to work with. It can also encourage platforms to improve their services and address any issues that are negatively affecting workers.
\item Advocate for better working conditions: Gig workers can come together and advocate for better working conditions, such as fair pay, better benefits, and protections. This can be done through social media campaigns, contacting policymakers, or even starting a union. By advocating for their rights, gig workers can help create positive change in the gig economy.
\item Stay informed about platform policies: Gig workers should regularly check the policies and terms of service of the platforms they use. This can help them understand their rights and responsibilities, as well as any changes that may affect their work. By staying informed, gig workers can avoid any negative surprises and make informed decisions.
\item Diversify income sources: Gig workers can reduce their reliance on a single gig platform by diversifying their income sources. This can be done by working with multiple platforms, finding offline work opportunities, or even starting their own business. By diversifying their income, gig workers can mitigate the risk of relying on a single platform and increase their overall earnings.
\end{enumerate}

\textbf {Summary of Problem-related Payment}\\
This category includes payment-related problems faced by gig workers on a gig work platform. These issues include fraudulent clients who don't pay, difficulties withdrawing funds, high fees, delayed payment, and issues with payment methods, such as exchange rates, tax information, and payment restrictions. Additionally, some freelancers are concerned about a change in policy that requires payment for connects or receiving a 1099-K form from the platform. Furthermore, some freelancers have trouble getting paid in their currency or complying with billing requirements. The lack of effective customer support and communication channels from the platform also adds to the difficulties. \\

\textbf{Solution suggested by LLM}\\

\begin{enumerate}
    
\item Communication is key: Gig workers should communicate with their clients before starting any project and clarify all the payment terms and conditions beforehand. They should ask for any external expenses and ensure that there are no surprises later on.
\item Use secure payment methods: Gig workers should use secure payment methods that are trusted and reliable. They should also be aware of the exchange rates and any fees associated with withdrawing money to avoid any issues later on.
\item Research the platform: Gig workers should research the gig work platform they intend to use and understand their fees and policies. They should also read reviews from other gig workers to ensure that they are using a reliable and trustworthy platform.
\item Keep track of earnings and tax information: Gig workers should keep track of their earnings and ensure that their tax information is up-to-date and accurate. They should also ensure that their bank account information matches the name on their gig work platform profile to avoid any issues with withdrawals.
\item Report fraudulent behavior: Gig workers should report any fraudulent behavior to the gig work platform's support team immediately. They should also keep all communication and evidence related to the project and payment to help the support team investigate and resolve the issue quickly.
\end{enumerate}

\textbf {Summary of Problem-related Poor Customer Support}\\
This category is about the poor customer support experienced by gig workers on the gig work platform. The problems include fraud clients who don't pay for work and unhelpful dispute teams, difficulties with withdrawing funds, errors in the account creation process, incorrect account bans, lack of access to live chat support, and unresponsive customer service. There are also problems with account reinstatement, delayed project approvals, and difficulty in logging in due to system errors. In addition, the high fees and long payment processing times, as well as the requirement to pay for Connects, are frustrating for freelancers. The customer support team is also unhelpful in addressing the concerns of freelancers whose gigs are denied, removed, or flagged as inappropriate. \\

\textbf{Solution suggested by LLM}\\

\begin{enumerate}

 \item  Protect yourself from fraud clients by always having a contract in place, with clearly defined milestones and deadlines. Use escrow payments or invoicing services to ensure that you receive payment for your work. In case of any dispute, provide clear evidence of the work done and communicate with the client politely, but firmly. If the dispute team is of no use, consider filing a complaint with the gig work platform or legal authorities.
 \item  If you face technical issues while working on the gig platform, such as login problems or error messages, reach out to customer support immediately. If you do not receive a response within a reasonable time, try different channels of communication, such as email, phone, or social media. Provide detailed information about the issue, including screenshots and error messages, and follow up regularly until the issue is resolved.
 \item If the live chat feature is not available or you are unable to get a response from customer support, try to connect with other gig workers on online forums or social media groups. They may have faced similar issues and can provide valuable advice and support.
 \item  If you are facing issues with reinstating your account or creating a new one, try to contact customer support through different channels and provide all the necessary information, such as your email address and phone number. If the support team is not helpful, consider switching to a different gig platform that has more user-friendly policies and procedures.
 \item  If you are unhappy with the fees charged by the gig platform or the quality of customer support, consider exploring other gig platforms that offer better rates or customer service. Read reviews and compare the features and policies of different platforms before making a decision. You can also join a professional association or community of gig workers to stay updated on industry trends and best practices.

\end{enumerate}

\textbf {Summary of Problem-related Scam}\\
Gig workers on gig work platforms face several scam-related problems, including fraud clients who don't pay, poor dispute resolution, and difficulty withdrawing funds. Other problems include receiving job offers outside the platform, downloading harmful files, fake job posts, unfair contract termination, and clients demanding high commissions after offering high wages. Furthermore, some clients use poor rating tactics to exploit workers, while others demand work at lower rates than the platform rate. Some clients also refuse to pay the required service fees and keep workers' earnings, while platforms charge excessive fees for services and tips. \\

\textbf{Solution suggested by LLM} \\

\begin{enumerate}
 \item Create a scam alert system - The gig platform can have an AI-based scam alert system that can track the behavior and patterns of the clients on the platform. The system can identify any suspicious behavior by analyzing the clients' work history, payment patterns, and other activities on the platform. If a client is identified as a potential scammer, the system can alert the gig workers before they start working with the client, saving them from any fraudulent activities. 
 \item Improved dispute resolution - Gig platforms should have a more efficient and responsive dispute resolution mechanism. A 24/7 support system with dedicated dispute resolution teams should be available to the gig workers. This will help them to resolve any payment or work-related issues quickly, ensuring that they are not scammed out of their hard work.
 \item  Introduce a verification system - Platforms can verify the client's identity and payment information to ensure that they are genuine. This can be done by verifying their phone numbers, email addresses, and other personal information. The platform can also ask for a deposit from the client, which can be refunded after the gig is completed. This will help to prevent any fraudulent activities and give the gig workers a sense of security.
 \item Educate gig workers about potential scams - Gig platforms should provide educational materials to gig workers on how to identify and avoid potential scams. These materials can include videos, articles, and webinars that cover various types of scams and how to avoid them. This will empower gig workers with knowledge and make them more aware of potential scams.
 \item  Encourage transparency - Gig platforms should encourage clients to be transparent in their job postings, payment processes, and communication with gig workers. Clients should disclose all the necessary details related to the job upfront, and payment processes should be clear and easy to understand. This will help gig workers to identify any red flags and avoid working with fraudulent clients.
 \end{enumerate}
\newpage
\subsection{\bf System Usability Scale Template}
Here are the questionaries for the System Usability Scale. It is measured on a scale of (1-5) where \textit{1=Strongly Disagree, 2=Disagree, 3=Neutral, 4=Agree, 5=Strongly Agree}. 
\begin{enumerate}
    \item  I think that I would like to use this system frequently.				
 \item I found the system unnecessarily
   complex.				
\item I thought the system was easy
   to use.                     	
\item I think that I would need the
   support of a technical person to
   be able to use this system.	
\item I found the various functions in
   this system were well-integrated.			
\item I thought there was too much
   inconsistency in this system.				
\item I would imagine that most people
   would learn to use this system
   very quickly.			
\item I found the system very
   cumbersome to use.			
\item I felt very confident using the
   system.	
\item I needed to learn a lot of
   things before I could get going
   with this system. 	
\end{enumerate}

After calculating the composite scores of SUS (System Usability Scale) scores, they are compared to the following adjectival rating table \cite{empiricalSUS}.
\begin{table}[ht]
\centering
\small 
\setlength{\tabcolsep}{0.1cm} 
\begin{tabular}{p{3cm} p{2cm} p{3cm} } 
SUS Score & Grade & Adjectival Rating \\ \midrule
>80.3 & A & Excellent \\
68-80.3 & B & Good \\
68 & C & Okay \\
51-68 & D & Awful \\
<51 & D & Poor
\end{tabular}
\\
\caption [Table showing adjectival ratings for SUS usability scores]{Adjectival Rating for SUS scores.}
\label{tab:SUS scores}
\end{table}

\end{document}